\documentclass[ preprint, 
 aps, pra ]{revtex4-1}
\usepackage{amsfonts}
\usepackage{amsmath}
\usepackage{graphicx}
\usepackage{dcolumn}
\usepackage{bm}

\begin{document}

\title{Exact Kohn-Sham Density Functional Theory on a Lattice}
\author{Kossi Amouzouvi}
\email{kossi@aims.edu.gh}
\author{Daniel P. Joubert}
\email{daniel.joubert2@wits.ac.za}
\date{\today}

\begin{abstract}
We formulate a set of equations that facilitate an exact numerical solution
of the Kohn-Sham potential for a finite Hubbard chain with nearest neighbour
hopping and arbitrary site potentials. The approach relies on a mapping of
the non-interacting Kohn-Sham ground state wave function onto the exact
interacting system wavefunction and two interconnected self-consistent
cycles. The self-consistent cycles are performed within the framework of the
Kohn-Sham non-interacting system without any direct reference to the
interacting system. The first self-consistent cycle updates the mapping of
the non-interacting wavefunction onto the interacting wavefunction based on
a trial input density, while the second self-consistent cycle updates the
Kohn-Sham potential to yield the trial density. At the solution point, the
exact density, the exact Kohn-Sham potential, the density functional
correlation energy and the exact interacting system ground state energy are
available.
\end{abstract}

\maketitle

\preprint{APS/123-QED}

\altaffiliation[Also at ]{AIMS Rwanda}

\affiliation{The National Institute for Theoretical Physics, School of
Physics and Mandelstam Institute for Theoretical Physics, University of the
Witwatersrand, Johannesburg, Wits 2050, South Africa}

%\pacs{Valid PACS appear here}





%\tableofcontents

%\section{Introduction}

Density Functional Theory (DFT) \cite{HohenbergKohn:64} on a Lattice (L-DFT)
has a history that dates back to its introduction by Sch\"{o}nhammer and
Gunnarson \cite{Schonhammer1987,Ijaes2010} in 1987. It is also called a Site
Occupation Number Function Theory (SOFT) \cite%
{Schoenhammer1995,Carrascal2015} since on a lattice, DFT is formulated in
terms of the site occupation numbers, based on the Hohenberg-Kohn theorem 
\cite{HohenbergKohn:64} that the ground state density is the controlling
variable which determines all the properties of the system. Kohn and Sham 
\cite{KohnSham:65} suggested an implementation of DFT where the interacting
systems is mapped onto a fictitious non-interacting system that yields the
same density as the interacting system. It is the Kohn-Sham (KS)
implementation of DFT that has lead to the popularity of DFT. The
fundamental principles on which DFT relies can be applied to any
many-particle Hamiltonian. Here we are interested in one dimensional Hubbard
Hamiltonian \cite{Hubbard1963,Lieb2003}. Despite its simplicity, the Hubbard
model \cite{Hubbard1963,Lieb2003} has been extremely successful in developing
an understanding of highly correlated many-particle systems. However, when
inhomogeneities are introduced, even simple lattice models quickly become
numerically intractable. Here L-DFT can play an important role \cite%
{PhysRevLett.56.1968,Schonhammer1987,Schoenhammer1995,Capelle2013}. Some of
the attractions of L-DFT is the simplicity of the model, the known solution
of the infinite uniform system \cite{Lieb1968} and the possibility to solve
the model for short finite chains exactly \cite%
{SWWB12,Jafari2008,Kingsley2013}. L-DFT has been used to explore formal
properties of exact functionals \cite%
{Campo2005,SWWB12,L.O.Wagner2012,Carrascal2015}. The model provides an ideal
environment for studying formal properties since it is possible to determine
exact properties, which is informative on their own, but it also provides a
reference to which approximations can be compared \cite%
{PhysRevA.79.032504,Franca2018}.

In this paper we present an exciting development for L-DFT. We have
formulated a set of equations that can be used iteratively to solve the
L-DFT problem for a finite chain exactly. Consider a finite one dimensional
single band interacting Hubbard Hamiltonian with nearest neighbour hopping 
\cite{Lieb2003}

\begin{equation}
\hat{H}=-t\sum_{<i,j>,\sigma }\hat{c}_{j\sigma }^{\dagger }\hat{c}_{i\sigma
}+U\sum_{i=1}\hat{n}_{i\uparrow }\hat{n}_{i\downarrow }+\sum_{i,\sigma }v_{i}%
\hat{n}_{i\sigma },  \label{def: Hubbard Hamiltonian}
\end{equation}%
where $\hat{c}_{i\sigma }^{\dagger }$ and $\hat{c}_{i\sigma }$ are the
creation and annihilation operators of an electron with spin $\sigma $ in
the localised state at site $i$, respectively. The summation $<i,j>$ is over
nearest neighbours only and we consider a finite system. $\hat{n}_{i\uparrow
}=$ $\hat{c}_{i\sigma }^{\dagger }\hat{c}_{i\sigma }$ is the site occupation
number operator for spin $\sigma ,$ while $v_{i}$ is a site potential and $U$
is the on-site interaction strength. We choose the hopping strength $t>0.$
The eigenspectrum of $\hat{H} $ is invariant under a change of the sign of $%
t,$ but the eigenfunctions are not \cite{Lieb2003}. The total spin operator $%
\sum_{i=1}\hat{n}_{i\sigma } $ commutes with the Hamiltonian,%
\begin{equation}
\left[ \sum_{i=1}\hat{n}_{i\sigma },\hat{H}\right] =0,  \label{spinC}
\end{equation}%
and therefore the numbers of up-spin, $N_{\uparrow },$ and down-spin, $%
N_{\downarrow },$ electrons are good quantum numbers. We are interested in
the lowest energy solutions of the Schr\"{o}dinger equation under the
constraint of a fixed number of up- and down-spin electrons, 
\begin{equation}
\hat{H}\left\vert \Psi (N_{\uparrow },N_{\downarrow })\right\rangle
=E\left\vert \Psi (N_{\uparrow },N_{\downarrow })\right\rangle .
\end{equation}%
As a consequence the spin constrained ground states of $\hat{H}$ can be
expended in terms of Slater determinants $\left\vert \phi _{m}\right\rangle $
with the same spin constraint as the ground state. The subscript $m\equiv
\left\{ N_{\uparrow },N_{\downarrow },N_{L}\right\} $ is a compound index
which represents all possible basis functions for a chosen spin
polarisation, $N_{\uparrow },$ $N_{\downarrow },$ and number of sites, $%
N_{L}.$ We follow the notation of Lieb and Wu \cite{Lieb2003} and label the
lattice sites sequentially from the left. The basis functions $\left\vert
\phi _{m}\right\rangle $ can be represented by an $n_{e}$-tuple ($n_{e}=$ $%
N_{\uparrow }+N_{\downarrow })$ 
\begin{eqnarray}
\left\vert \phi _{m}\right\rangle &=&\hat{c}_{x_{1}^{m}\uparrow }^{\dagger }%
\hat{c}_{x_{2}^{m}\uparrow }^{\dagger }\cdots \hat{c}_{x_{N_{\uparrow
}}^{m}\uparrow }^{\dagger }\hat{c}_{x_{N_{\uparrow }+1}^{m}\downarrow
}^{\dagger }\hat{c}_{x_{N_{\uparrow }+2}^{m}\downarrow }^{\dagger }\cdots 
\hat{c}_{x_{n_{e}}^{m}\downarrow }^{\dagger }|\rangle  \notag \\
&\equiv &\left( x_{1}^{m},x_{2}^{m},....,x_{n_{e}}^{m}\right)
\label{def: Phi m}
\end{eqnarray}%
where the first $N_{\uparrow }$ entries, $x_{i}^{m},$ $%
x_{i}^{m}<x_{i+1}^{m}, $ label the sites $x_{i}^{m}$ in the Slater
determinant with $\left\vert \phi _{m}\right\rangle $ with occupied up-spin
states and the $N_{\uparrow }+1$ to $n_{e}$ entries, $x_{i}^{m}<x_{i+1}^{m},$
label the sites where localised down-spin states are occupied . The states $%
\left\vert \Psi (N_{\uparrow },N_{\downarrow },v)\right\rangle $ can be
expanded in terms of the corresponding spin constrained basis functions as 
\begin{equation}
\left\vert \Psi (N_{\uparrow },N_{\downarrow },v)\right\rangle =\sum
f_{m}\left\vert \phi _{m}\right\rangle .  \label{expand}
\end{equation}

The lattice non-interacting Hamiltonian $\hat{H}^{0}$ () is defined as 
\begin{equation}
\hat{H}^{0}\left( v^{0}\right) =-t\sum_{<i,j>,\sigma }\hat{c}_{j\sigma
}^{\dagger }\hat{c}_{i\sigma }+\sum_{i,\sigma }v_{i}^{0}\hat{n}_{i\sigma },
\label{Hdft}
\end{equation}%
with ground state solutions 
\begin{equation}
\left\vert \Phi (N_{\uparrow },N_{\downarrow },v^{0})\right\rangle =\sum
g_{m}\left\vert \phi _{m}\right\rangle .  \label{expand_dft}
\end{equation}%
When the site potential $v^{0}$ is equal the Kohn-Sham site potential, $%
v^{KS},$ the interacting and non-interacting site densities, $n_{i},$ are
identical,%
\begin{equation}
n_{i}=\left\langle \Psi \right\vert \hat{n}_{i\uparrow }+\hat{n}%
_{i\downarrow }\left\vert \Psi \right\rangle =\left\langle \Phi \right\vert 
\hat{n}_{i\uparrow }+\hat{n}_{i\downarrow }\left\vert \Phi \right\rangle
\label{site_densities}
\end{equation}%
For ease of notation, we have suppressed the total spin and potential in the
notation of the wave functions.

We partition the interacting system energy as%
\begin{equation}
E=\left\langle \Psi \left\vert \hat{H}\right\vert \Psi \right\rangle
=T^{KS}+E^{hx}+E^{c}+\left\langle \Phi \left\vert \hat{v}\right\vert \Phi
\right\rangle
\end{equation}%
where 
\begin{eqnarray}
T^{KS} &=&\left\langle \Phi \left\vert \hat{T}\right\vert \Phi \right\rangle
\notag \\
E^{hx} &=&\left\langle \Phi \left\vert \hat{u}\right\vert \Phi \right\rangle
\notag \\
E^{c} &=&\left\langle \Psi \left\vert \hat{T}+\hat{u}\right\vert \Psi
\right\rangle -\left\langle \Phi \left\vert \hat{T}+\hat{u}\right\vert \Phi
\right\rangle  \notag \\
\left\langle \Phi \left\vert \hat{v}\right\vert \Phi \right\rangle
&=&\left\langle \Psi \left\vert \hat{v}\right\vert \Psi \right\rangle
=\sum_{i}v_{i}n_{i}  \label{hxc}
\end{eqnarray}%
$E^{c}$ is has the form of the conventional correlation energy while $E^{hx}$
is the sum of the equivalent conventional Hartree and exchange energies. The
ground state energies are functions of the site densities \cite%
{Schoenhammer1995} and the requirement that the ground state site densities
of the interacting and non-interacting systems are identical requires that 
\cite{KohnSham:65} 
\begin{equation}
v_{i}^{KS}=v_{i}+\frac{d}{dn_{i}}\left( E^{hx}+E^{c}\right) .
\end{equation}%
The hx term in Eq.(\ref{hxc}), by definition, depends on the Kohn-Sham (KS)
ground state wavefunction only. When the correlation is set to zero, the KS
equation can be solved self-consistently without further approximation. The
correlation term, on the other hand, depends on the interacting and
non-interacting system ground state wavefunctions and is the term that has
proven to be the most challenging to approximate. Here we follow a different
approach and formulate an expression for the correlation energy that can be
solved within a non-interacting framework without any direct reference to
the interacting system.

The interacting and non-interacting wavefunctions can be expanded in terms
of the same basis functions and we map the non-interacting wave function
onto the interacting wavefunction 
\begin{equation}
\left\vert \Psi \right\rangle =\hat{J}\left\vert \Phi \right\rangle .
\end{equation}

with 
\begin{equation}
\hat{J}=\sum_{m}e^{-\gamma _{m}{\ \hat{x}^{m}}}
\end{equation}%
where the projection operator {\ $\hat{x}^{m}=\hat{x}_{1}^{m}\cdots \hat{x}%
_{n_{e}^{m}}$ }The $\hat{x}_{i}$ are equal to $\hat{n}_{x_{i}\uparrow }$ if $%
i\leq N_{\uparrow }$ and $\hat{n}_{x_{i}\downarrow }$ otherwise. The
expansion coefficients in Eqs. (\ref{expand}) and (\ref{expand_dft}) are
simply related, 
\begin{equation}
f_{m}=e^{-\gamma _{m}}g_{m}.  \label{map}
\end{equation}%
We can work in real space without losing generality. For the single band
linear Hubbard model with nearest neighbour mapping, the ground state
wavefunction is non-degenerate and for $t>0,$ the expansion coefficients $%
f_{m}$ and $g_{m}$ are non-zero and all have the same sign \cite{Lieb2003}.
It follows that when the $f_{m}$ and $g_{m}$ are chosen to have the same
sign, the mapping in Eq. (\ref{map}) is well defined. The operator $\hat{J}$
commutes with $\hat{u}$ and $\hat{v}$ and we can write%
\begin{equation}
{\ \left( \hat{J}^{-1}\hat{T}\hat{J}+\hat{u}+\hat{v}\right) \left\vert \Phi
\right\rangle =E\left\vert \Phi \right\rangle }
\end{equation}%
where $E$ is the ground state energy of the interacting system. Taking the
expectation value with respect to $\left\vert \Phi \right\rangle ,$ and
assuming that all wavefunctions are normalised , the interacting system
energy can be expressed as 
\begin{equation}
\left\langle \Phi \right\vert {\ \hat{J}^{-1}\hat{T}\hat{J}+\hat{u}+\hat{v}%
\left\vert \Phi \right\rangle =E.}  \label{EInt}
\end{equation}%
Knowledge of ${\ \hat{J}}$ is then sufficient to find the interacting energy
with reference to the non-interacting system ground state wavefunction only.

We have

{\ 
\begin{equation}
\langle \phi _{m}|\hat{J}^{-1}\hat{T}\hat{J}\Phi \rangle
=-t\sum_{i=1}^{n_{e}}\sum_{s=\pm 1}g(x^{m}[x_{i}^{m}+s])e^{-\gamma
_{x^{m}[x_{i}^{m}+s]}}e^{\gamma _{m}}  \label{gm}
\end{equation}%
}where $x^{m}[x_{i}^{m}\pm 1]$ represents the $n_{e}$-tuple with the
occupancy of site $x_{i}^{m}$ decreased by one and the occupancy of a
neighbouring site (+ to the right and - to the left) increased by one. If
this results in a double occupancy of the same spin or if a non-existing
site of the finite chain is occupied, the resulting coefficient $%
g(x^{m}[x_{i}^{m}\pm 1])=0.$ From Eq. (\ref{hxc}) the correlation energy can
be written as 
\begin{eqnarray}
E^{c} &=&\left\langle \Phi \right\vert \hat{J}^{-1}\hat{T}\hat{J}-\hat{T}%
\left\vert \Phi \right\rangle   \notag \\
&=&-t\sum_{m,m^{\prime }}g\left( x^{m}\right) \sum_{i=1}^{n_{e}}\sum_{s=\pm
1}g\left( (x^{m}\left[ x_{i}^{m\prime }+s\right] \equiv x^{m}\right) \left(
e^{-\gamma _{x^{m\prime }}}e^{\gamma _{x^{m}}}-1\right) ,  \label{E_c}
\end{eqnarray}%
and from Eqs. (\ref{EInt}) and (\ref{gm}) we get {\ \ }%
\begin{equation}
{e^{\gamma _{m}}=\dfrac{g(x^{m})\langle \phi _{m}|(E-\hat{u}-\hat{v})\phi
_{m}\rangle }{-t\sum_{i=1}^{n_{e}}\sum_{s=\pm
1}g(x^{m}[x_{i}^{m}+s])e^{-\gamma _{x^{m}[x_{i}^{m}+s]}}}.}  \label{gamma}
\end{equation}%
The expressions in Eqs. (\ref{E_c}) and (\ref{gamma}) are correct at the
solution point only. The site densities for the interacting, $n_{i},$ and
non-interacting, $n_{i}^{0},$ systems can be expressed as{\ }

\begin{equation}
n_{i}=\sum_{m}e^{-2\gamma _{m}}g^{2}\left( x^{m}\right) \left\langle
x^{m}\left\vert \hat{n}_{i\uparrow }+\hat{n}_{i\downarrow }\right\vert
x^{m}\right\rangle
\end{equation}%
and 
\begin{equation}
n_{i}^{0}=\sum_{n}g^{2}\left( x^{m}\right) \left\langle x^{m}\left\vert \hat{%
n}_{i\uparrow }+\hat{n}_{i\downarrow }\right\vert x^{m}\right\rangle .
\end{equation}

In the Kohn-Sham formalism, the site densities of the interacting and
fictitious KS-system must be identical, $n_{i}=n_{i}^{0},$ and this is key
to the solution outlined below. It is possible to derive expressions for the
Kohn-Sham site potentials $v_{i}+\frac{d}{dn_{i}}\left( E^{hx}+E^{c}\right) $
using the site analogue of the outline given in \cite{SMVG:97,SMMVG:99}, but 
$E$ and $\left\vert \Psi \right\rangle =\hat{J}\left\vert \Phi \right\rangle 
$ can be found without knowledge of $\frac{d}{dn_{i}}\left(
E^{hx}+E^{c}\right) $ by directly determining $v_{i}^{KS}$ iteratively in a
double self-consistent calculation as follows:

Approximate $v_{i}^{0}$ and find the ground state wavefunction for the
corresponding non-interacting Hamiltonian $H^{0},$ of Eq. (\ref{Hdft}).
Choose an initial set of parameters $\left\{ \gamma _{m}\right\} .$ Repeat
the following self-consistent cycles till self consistency is reached:

\begin{enumerate}
\item Use Eq. (\ref{EInt}) to approximate $E\left\{ \gamma _{m}\right\} .$
Iterate Eq. (\ref{gamma}) for $\left\{ \gamma _{m}\right\} $ to a
self-consistent solution for the fixed input $g(x^{m})$ from the solution of 
$H^{0}.$

\item Determine $n_{i}\left\{ \gamma _{m}\right\} $ from previous solution
of $g\left( x^{m}\right) $ and $\left\{ \gamma _{m}\right\} $ from cycle 1.
Keep $n_{i}\left\{ \gamma _{m}\right\} $ fixed and update $v_{i}^{0}$ till $%
n_{i}^{0}\left[ v^{0}\right] =n_{i}\left\{ \gamma _{m}\right\} .$ Use the
output $g\left( x^{m}\right) $ and repeat cycle 1.
\end{enumerate}

When the two cycles reach self-consistency simultaneously, we have a
solution where the interacting and non-interacting densities are identical.
This is simply the requirement for, or definition of, the KS fictitious
system, the non-interacting system that gives the same density as the fully
interacting system \cite{KohnSham:65}. Therefore, at self-consistency, $%
v^{0}=v^{KS}.$ Note that if Eq. (\ref{gamma}) is satisfied, the energy $%
E\left\{ \gamma _{m}\right\} $ is the variationally minimised energy of $H$
under the constraint that the expansion coefficients $e^{-\gamma
_{m}}g\left( x^{m}\right) $ all have the same sign. Since we use a complete
basis set for each spin polarisation, $E\left\{ \gamma _{m}\right\} $ is the
interacting ground state energy for the corresponding spin polarisation. In
this approach we not only get the exact KS-lattice potential, but also the
exact interacting particle wave function.

If is a simple matter to perform the diagonalisation of the KS-system by
determining the eigensolutions of a single particle Hamiltonian and then
expand the many-particle ground state wavefunction in terms of the
many-particle basis functions introduced above. This makes the approach
flexible and allows solution for long finite chains.

In our tests for Hubbard chains up to 10 sites long, a simple
self-consistent iteration cycle, where a fraction of the output of a cycle, $%
\left\{ \gamma _{m}\right\}$ or $\left\{ v_{i}^{0}\right\}$, is added to the
corresponding value of the previous cycle, scaled by an appropriate factor,
always converged to the exact value.

In conclusion, we have formulated a set of equations that allows us to solve
the Kohn-Sham equations exactly for a finite Hubbard chain with nearest
neighbour hopping. The approach relies on the unique property of the model
that the expansion coefficients of the ground state wavefunction in terms of
an appropriate set of Slater determinant basis functions have the same sign.
The ideas introduced may, however, have potential in a more general setting.

\section*{Acknowledgements}

DPJ acknowledges support from the National Research Foundation (NRF) and KA
acknowledges supports from the Deutscher Akademischer Austauschdienst (DAAD)
and AIMS Rwanda.

\bibliographystyle{apsrev}
\bibliography{c:/journals/bibtex/dft}

\end{document}